\documentclass[10pt,showpacs,amsmath,amssymb]{article}
\usepackage{amsfonts}
\usepackage{mathdots}
\usepackage{color}

\newcommand{\field}[1]{\mathbb{#1}}

\begin{document}

\title{Constant curvature solutions of Grassmannian sigma models: (1) Holomorphic solutions}

\author{L. Delisle${}^{1,4}$, V. Hussin${}^{1,2,5}$ and W. J. Zakrzewski${}^{3,6}$}

\footnotetext[1]{D\'epartement de Math\'ematiques et de
Statistique, Universit\'e de Montr\'eal, C.P. 6128,
Succ.~Centre-ville, Montr\'eal (Qu\'ebec) H3C 3J7, Canada.}
\footnotetext[2]{Centre de Recherches Math\'ematiques,
Universit\'e de Montr\'eal, C.P. 6128, Succ.~Centre-ville,
Montr\'eal (Qu\'ebec) H3C 3J7, Canada.}
\footnotetext[3]{Department of Mathematical Sciences, University of Durham, Durham DH1 3LE, United Kingdom.}
\footnotetext[4]{email:delisle@dms.umontreal.ca}
\footnotetext[5]{email:hussin@dms.umontreal.ca}
\footnotetext[6]{email:w.j.zakrzewski@durham.ac.uk}

\date{\today}

\maketitle

\begin{abstract}
We present a general formula for the Gaussian curvature of curved holomorphic 
2-spheres in Grassmannian manifolds $G(m,n)$. We then show how to
 construct such solutions with constant curvature. We also make some relevant conjectures for the admissible constant curvatures in  $G(m,n)$ and give some explicit expressions, in particular, for  $G(2,4)$ and $G(2,5)$.  
\end{abstract}



\section{Introduction \label{intro}}

Just over 10 years ago Li and Yu discussed, in a very interesting 
paper${}^{1}$, the classification of minimal 2-spheres with constant Gaussian curvatures (called 'curvature' in this paper) immersed in the complex Grassmannian manifold $G(2,4)$.

The classification of Li and Yu came in the form of two theorems (named $A$ and $B$ in their paper).
Theorem $A$ discussed the holomorphic case and it showed that only four constant curvatures
were possible. These are given as $K=4,2,\frac{4}{3}$ and $1$. Moreover,  their paper also gave explicit examples of such holomorphic immersions. Theorem $B$ described non-holomorphic immersions which correspond to $K=2,1,\frac{2}{3}$ and $\frac{2}{5}$.

Since the appearance of this paper many other papers have also been written. In particular, a classification of holomorphic spheres with constant curvature has been 
produced ${}^{2}$ for $G(2,5)$. Other contributions by the same authors mostly dealt with non-holomorphic immersions${}^{3,4,5}$ for $G(2,n)$.

Having recently looked${}^{6}$ at the construction of higher-dimensional surfaces based on harmonic maps of $S^2$ into $\field{C}P^{n-1}$ and other Grasmannian manifolds we have tried to understand the theorems of Li and Yu using our approach. The non-holomorphic maps (Theorem B)  were relatively easy to understand by taking, for example, pairs of maps${}^{6}$ from the Veronese sequence. This, however,  was not the case for the holomorphic maps (Theorem A).  

Motivated by this task we have started thinking about the explicit expressions for the curvatures of the holomorphic immersions. This has lead to the expressions given in the next section and they, in turn, provide us with an explicit algorithmic procedure for the construction of all constant curvature immersions in complex Grassmannian manifolds $G(m,n)$. When we apply this procedure to $G(2,4)$ we recover the results${}^{1}$ of Li and Yu. We have also retrieved, in a way that will be clarified in the paper, the results of Jiao and Peng${}^{2}$.

In Section 2, we present the general formulae for the  curvature of holomorphic immersions into $G(m,n)$. Two different parametrizations of the solutions are introduced. In Section 3, we discuss such immersions for the case of constant  curvature. Some of them are related to the  Veronese curves in $G(m,n)$ and we conjecture that the Veronese curves give rise to the smallest possible curvatures. All admissible curvatures for holomorphic solutions are then discussed and we make a second conjecture. Section 4 is devoted to the case of  $G(2,n)$ where our conjectures are partially proved and where the cases $G(2,4)$ and $G(2,5)$ are discussed in detail.

\section{Energy density and  curvature of holomorphic immersions of 2-spheres in Grassmannian manifolds}
\subsection{General discussion}

Let us consider maps of $S^2$ into a Grassmaniann manifold $G(m,n), \ n>m$. One way of thinking about  such maps involves
treating them as fields of $U(m,n)$ $\sigma$ models${}^{7}$.
Thus we consider $n\times m$ complex fields put in an array $Z=\{Z_{\alpha,a}, \alpha=1,2...n$, $a=1,..,m\}$ (note the order of indices).
These fields are subject to the constraint
\begin{equation}
Z^{\dagger}Z\,={\mathbb I}_m,
\label{const}
\end{equation}
where ${\mathbb I}_m$ is the $m\times m$ unit matrix.
The fields transform as
\begin{equation}
Z\,\rightarrow Z'=VZU
\label{trans}
\end{equation}
under global $V\in U(n)$ and local $U\in U(m)$ transformations.

Then one introduces  $U(m)$ matrix valued gauge fields $A_{\mu}$, which
transform as
\begin{equation}
A_{\mu}\,\rightarrow \,U^{\dagger}A_{\mu}U\,-\,iU^{\dagger}\partial_{\mu}U
\label{gauge}
\end{equation}
and uses them to construct gauge covariant derivatives
\begin{equation}
D_{\mu}Z\,=\,\partial_{\mu}Z\,-\,iZA_{\mu}.
\label{covariant}
\end{equation}
Here $\mu$ stands for the index of  $x_1$ and $x_2$ - two local coordinates on the Euclidean space $\field{R}^2$.

Then keeping the constraint (\ref{const}) and defining the gauge fields in terms of $Z$ and its derivatives
\begin{equation}
iA_{\mu}\,=\,Z^{\dagger}\partial_{\mu}Z,
\label{gauge2}
\end{equation}
we define the Lagrangian density (``energy" - in the harmonic map literature) as
\begin{equation}
{\cal L}\,=\,\frac{1}{4}\hbox{Tr}  (D_{\mu}Z)^{\dagger} D_{\mu}Z.
\label{lagrangian}
\end{equation}
The invariance of the corresponding action $S=4\int_{\Omega}{\cal L} dx_1 dx_2$ leads to the Euler-Lagrange equations 
\begin{equation}
D_{\mu} D_{\mu} Z+ Z {(D_{\mu}Z)^{\dagger}}(D_{\mu}Z)=0,
\label{eulerlagrange}
\end{equation}
where $\Omega$ is an open, connected subset of the Euclidean space $\field{R}^2$ and $x_{1}$ and $x_{2}$ are local coordinates on $\Omega$. The model described by
(\ref{lagrangian}) and (\ref{eulerlagrange}) is the so-called two-dimensional $U(m,n)$ Grasmannian $\sigma$ model.

Next we introduce the complex variables and corresponding derivatives in our Euclidean space $\field{R}^2$
by defining 
\begin{equation}
x_{\pm}\,=\,x_1\pm ix_2, \quad  \partial\,=\,\frac{1}{2}(\partial_{x_1}-i
\partial_{x_2}), \bar\partial\,=\,\frac{1}{2}(\partial_{x_1}+i
\partial_{x_2}).
\label{holo}
\end{equation}

In these coordinates, the Lagrangian density is given as
\begin{equation}
{\cal L}\,=\,\frac{1}{2}\hbox{Tr}\left[(DZ)^{\dagger}DZ+({\bar D}Z)^{\dagger}{\bar D}Z\right],\label{lagrangiancomplex}
\end{equation}
and the equations of motion are
\begin{equation}
 {\bar D}DZ+Z(DZ)^{\dagger}DZ=0.\label{eulerlagrangecomplex}
\end{equation}

Using the invariance properties of the theory, we can introduce two different parametrizations  of the matrix field $Z$. The first one ${}^{8}$, called the Macfarlane parametrization, is given as 
\begin{equation}
Z\,=\hat Z \hat L\,=\,\left(\begin{array}{c}
{\mathbb I}_m \\
K
\end{array}\right)\hat L\ ,
\label{part1}
\end{equation}
where $K$ is a $(n-m)\times m$ matrix and $L$ is a $m\times m$ matrix,
which due to (\ref{const}) satisfy
\begin{equation}
{\hat L}^{\dagger} (K^{\dagger}K+{\mathbb I}_m) {\hat L}\,=\,{\mathbb I}_m \quad {\rm or} \quad
 K^{\dagger}K +{\mathbb I}_m \,=\,({\hat L} {\hat L}^{\dagger} )^{-1}.
\label{para}
\end{equation}
Since the $m\times m$ matrix $({\hat L} {\hat L}^{\dagger})$ is clearly  Hermitian and strictly positive definite, its inverse exists and we define it as $M$ such that
\begin{equation}
\hat M \,=\,({\hat L} {\hat L}^{\dagger} )^{-1}\,=\,{\mathbb I}_m+ K^{\dagger}K={\hat Z}^{\dagger} \hat Z.
\label{em}
\end{equation}
Then${}^{8}$, if $K$ is holomorphic, {\it i.e.} $\bar \partial K=0$ the corresponding $Z$ satisfies the Euler-Lagrange equations (\ref{eulerlagrangecomplex}) for the Lagrangian (\ref{lagrangiancomplex}). They correspond to the holomorphic immersions of $S^2$ into the Grasmannian manifold $G(m,n)$.

The second parametrization${}^{7}$, called orthogonalized parametrization, is given by  
\begin{equation}
Z\,=\, \tilde Z \tilde L,
\label{moreh}
\end{equation}
where $\tilde Z$ is a holomorphic $n\times m$ matrix obtained from a set a linearly independent holomorphic vectors $f_1,\dots,f_m$ and $\tilde L$ is a $m\times m$ matrix . Due to (\ref{const}) 
we  obtain
\begin{equation}
{\tilde M}=({\tilde L} {\tilde L}^{\dagger})^{-1}={\tilde Z}^{\dagger}{\tilde Z},
\label{em1}
\end{equation}
a relation which is similar to (\ref{em}).

\subsection{Energy density and  curvature }

As holomorphic immersions satisfy $\bar\partial \hat Z=0$ (or $\bar \partial K=0$),
for them the Lagrangian (\ref{lagrangiancomplex}) takes the form
\begin{equation}
{\cal L}\,=\,\frac{1}{2}\hbox{Tr}  (DZ)^{\dagger} DZ.
\label{lagrangiana}
\end{equation}

Let  us now use the Macfarlane parametrisation (\ref{part1}) to calculate (\ref{lagrangiana}).
First we look at the case of the $\field{C}P^{n-1}$ fields in which $\hat Z$ is a $n$ component 
vector. In this case, we have
\begin{equation}
{\cal L}\,=\, \frac{1}{2}\left(\frac{\vert \partial {\hat Z}\vert^2}{\vert {\hat Z}\vert^2}-\frac{\vert {\hat Z}^{\dagger}\partial {\hat Z}\vert^2}{\vert {\hat Z}\vert^4} \right).
\label{exp1}
\end{equation}
However, given that $\bar\partial \hat Z=0$  (and so $\partial \hat Z^{\dagger}=0$)
this can be rewritten as
\begin{equation}
{\cal L}\,=\,\frac{1}{2} \left( \frac{\partial \bar\partial \vert \hat Z\vert^2}{\vert \hat Z\vert^2}
 \,-\,\frac{\bar \partial \vert \hat Z\vert^2 \,\partial \vert \hat Z\vert^2}{\vert \hat Z\vert ^4}\right)
\label{cp1}
\end{equation}
and finally
\begin{equation}
{\cal L}\,=\,\frac{1}{2} \partial\left(\frac{\bar{\partial}\vert \hat Z\vert^2}{\vert \hat Z\vert^2}\right)
\,=\,\frac{1}{4} \partial\left(\bar\partial \,\ln \vert \hat Z\vert^2\right).
\label{cpa}
\end{equation}

We consider now the general Grassmannian manifold $G(m,n)$ described by the field (\ref{part1}).
It is easy to check that the two terms in (\ref{exp1}) are replaced by (up to an overall factor 
$\frac{1}{2}$)
\begin{equation}
\hbox{Tr}\,\left(\bar \partial \hat Z^{\dagger}\partial \hat Z {\hat M}^{-1}\,-\,
(\partial \hat Z)^{\dagger} \hat Z {\hat M}^{-1} {\hat Z}^{\dagger} (\partial \hat Z) {\hat M}^{-1}\right),
\label{terms}
\end{equation}
where ${\hat M}$ is given by (\ref{em}).
However, using the property that $\partial {\hat Z}^{\dagger}=\bar\partial {\hat Z}=0$ we rewrite (\ref{terms}) as
\begin{equation}
\,\partial \hbox{Tr}\,({\hat M}^{-1}\,\bar\partial {\hat M}).
\label{gena}
\end{equation}
As expected, in the $\field{C}P^{n-1}$ case, this reduces to (\ref{cpa}).

Next we show that (\ref{gena}) is given by $\partial \bar \partial \,
 \hbox{ln} \det {\hat M}$.  Indeed, since ${\hat M}^{-1}=(\det {\hat M})^{-1}adj({\hat M})$, we obtain
 \begin{equation}
 \hbox{Tr}({\hat M}^{-1}\,\bar\partial {\hat M})=\frac{1}{det {\hat M}}\sum_{i,j=1}^m \, adj({\hat M})_{ij}\bar \partial {\hat M}_{ji}.
 \end{equation}
But $\hbox{Tr} ({\hat M}\,adj({\hat M})\,=\,m \det{{\hat M}}$ and its easy to check that (for $m>1$)
$\hbox{Tr} ((\bar \partial adj({\hat M}) {\hat M})=(m-1)\hbox{Tr} ( adj({\hat M}) \bar \partial {\hat M})$
(as all the minors in $adj({\hat M})$ involve products of $m-1$ terms).

Finally, putting everything together we see that for the holomorphic embeddings the Lagrangian becomes
\begin{equation}
{\cal L}\,=\frac{1}{2} \,\partial \bar\partial \,\hbox{ln}\det {\hat M}\,
\label{Lfinal}
\end{equation}
and the associated  curvature is given as
\begin{equation}
{\cal K}\,= -\frac{1}{{\cal L}}\, \partial \bar\partial \,\hbox{ln}\, {\cal L}.
\label{expcurv}
\end{equation}

Here, it is worth noticing that the type of parametrization of $Z$, ({\it i.e.} (\ref{part1}) or (\ref{moreh})), does not affect the proof of this result since the only assumption 
used in its derivations has been the fact that $\hat Z$ or $\tilde Z$ is holomorphic. In the case (\ref{part1}), the Lagrangian is given by (\ref{Lfinal}) with
\begin{equation}
\det {\hat M}\,=\, \vert \det \hat L \vert^{-2}\,=\, \det(\mathbb{I}_m + K^\dagger K),
\label{exp}
\end{equation}
while for the case  (\ref{moreh}), the Lagrangian is given by 
\begin{equation}
{\cal L}\,=\frac{1}{2} \,\partial \bar\partial \,\hbox{ln}\det {\tilde M}\,
\label{Lfinal2}
\end{equation}
with
\begin{equation}
\det {\tilde M}\,=\, \vert \det {\tilde L} \vert^{-2}.
\label{exp}
\end{equation}

\subsection{Orthogonalized parametrization and a wedge product }
Let us here explain the choice of the parametrization (\ref{moreh}) and its name.
In ${\tilde M}$, the matrix ${\tilde L}$ may be chosen triangular since we start from holomorphic linearly independent vectors $f_1,\dots,f_m$  that we orthonormalize to obtain $Z$. Indeed, the matrix ${\tilde L}=\{ l_{ij}, i,j=1,\ldots, m\}$ is thus given as
\begin{eqnarray}
\left\{
\begin{array}{cc}
l_{ii} =  \frac{1}{\vert {\tilde f_i}\vert}\,, & \quad
i = 1, \ldots,m\, ,\\
l_{ij} =- \frac{1}{\vert {\tilde f_j}\vert}\frac{f_i^\dagger  f_j}{\vert {\tilde f_i}\vert^2}\,, &
 \qquad i<j\,, \\
 l_{ij} = 0\,, & \quad {\rm if} \,\,\,\,i>j,
\end{array}\right.
\label{amatrix}
\end{eqnarray}
where the vectors $\tilde f_2,\dots,\tilde f_m$ are obtained from $f_1,\dots,f_m$ by the well-known Gram-Schmidt orthogonalization process. We thus obtain 
\begin {equation}
\det {\tilde M}\,=\, \vert \det {\tilde L} \vert^{-2}= \vert { f_1}\vert^2 \prod_{i=2}^{m}\vert {\tilde f_i}\vert^2.
\label{detem1}
\end{equation}

In particular, for $G(2,n)$, we see that we have
\begin{equation}
{\tilde M}\,=\, \left(\begin{array}{cc}
\vert f_1\vert^2 &  f_1^\dagger f_2 \\
f_2^\dagger f_1 & \vert f_2\vert^2
\end{array}\right),
\label{gras2}
\end{equation}
which can be diagonalized as
\begin{equation}
\tilde M_D\,=\,\left(\begin{array}{cc}
\vert f_1\vert^2 &  0 \\
0 & \vert \tilde f_2\vert^2
\end{array}\right),
\label{gras2}
\end{equation}
with 
\begin{equation}
 \label{othro2}
 \tilde f_2\,=\, f_2\,-\, \frac{f_1^{\dagger} f_2}{\vert f_1\vert^2} f_1.
\end{equation}
We thus obtain
\begin {equation}
\det {\tilde M} \,= \vert { f_1}\vert^2 \vert {\tilde f_2}\vert^2=  \vert { f_1}\vert^2 \vert { f_2}\vert^2- \vert { f_1}^{\dagger}  f_2 \vert^2.
\end{equation}

Note that we can also define a `wedge product' of two vectors; namely,
$f_1\wedge f_2$.
Such a quantity gives us an $n \times n$ matrix defined as
\begin{equation}
 \label{wedge1}
(f_1\wedge f_2)_{ij}\,=\, {\cal A}_{ij}\,=\, (f_1)_i(f_2)_j\,-\, (f_2)_i(f_1)_j.
\end{equation}
Then it is easy to check that
\begin{equation}
 \label{wedge2}
\vert f_1\wedge f_2\vert^2\,=\, 
\sum_{ij}  {\cal A}_{ij}^{\dagger}  {\cal A}_{ij} \,=\, 2(\vert f_1\vert ^2\vert f_2\vert ^2\,-\,
\vert f_1^{\dagger}f_2\vert^2)\, =\, 2
\det {\tilde M}.
\end{equation}
This suggests another way of thinking of $\det {\tilde M}$ which can be generalized to $G(m,n)$.

Indeed, we can write
\begin{equation}
{\cal L}\,=\, \frac{1}{2}\,\partial \bar \partial \,\ln\, {\cal A}^{\dagger} {\cal A},
\label{a}
\end{equation}
with 
\begin{equation}
  {\cal A}^{\dagger} {\cal A}\,=\,\sum_{i_1,i_2,\cdots i_m} \, \vert  {\cal A}_{i_1,i_2,\cdots i_m}\vert ^2=\, m!\,
\det {\tilde M}.
\label{b}
\end{equation}
and where 
\begin{equation}
\label{c}
  {\cal A}_{i_1,i_2,\cdots i_m}\,=\, (f_1\wedge f_2\wedge\cdots \wedge f_m)_{i_1,i_2,\cdots i_m}.
  \label{Agen}
\end{equation}

\subsection{Macfarlane parametrization and general results}

Since the Grassmannian manifold $G(m,n)$ is defined as the coset space $ \frac{U(n)}{U(m)\times U(n-m)}$, we have the well-known duality property $G(m,n)\cong G(n-m,n)$. We can thus use the Macfarlane parametrization to prove the following theorem:

\medskip

\textbf{Theorem 1:} The holomorphic solutions of $G(m,n)$ are in one-to-one correspondence with the holomorphic solutions of $G(n-m,n)$. Furthermore, we have  
\begin{equation}
 \det {\hat M}^{(m,n)}=\det {\hat M}^{(n-m,n)}.\label{equathm1}
\end{equation}
\textit{Proof:} Let
\begin{equation}
\varphi:\ G(m,n)\ \rightarrow \ G(n-m,n)\label{bijec1}
\end{equation}
be defined by
\begin{equation}
 \varphi(\hat Z^{(m,n)})=\varphi\left(\left(\begin{array}{c}
{\mathbb I}_m \\
K^{(m,n)}
\end{array}\right)\right)=\left(\begin{array}{c}
{\mathbb I}_{n-m} \\
K^{(m,n)^{T}}
\end{array}\right)=\hat Z^{(n-m,n)}.
\label{bijec2}
\end{equation}
We see that, since the transposition map is one-to-one, the map $\varphi$ is one-to-one and $\varphi$ establishes a one-to-one correspondence between the Macfarlane parametrization of $G(m,n)$ and the Macfarlane parametrization of $G(n-m,m)$. To prove the second part of the theorem we calculate $\det {\hat M}^{(n-m,n)}$. We find 
\begin{eqnarray}
 \det {\hat M}^{(n-m,n)}&=&\det\left({\mathbb I}_{n-m}+(K^{(m,n)^{T}})^{\dagger}K^{(m,n)^{T}}\right)\\
&=&\det\left({\mathbb I}_{n-m}+K^{(m,n)}(K^{(m,n)})^{\dagger}\right)\\
&=&\det\left({\mathbb I}_{m}+(K^{(m,n)})^{\dagger}K^{(m,n)}\right)\\
&=&\det {\hat M}^{(m,n)}.
\end{eqnarray}
The last equality  is a direct consequence of the Sylvester's determinant theorem. This concludes the proof.

From Theorem 1, we note that if ${\cal K}$ is the curvature (not necessarily constant) of a holomorphic solution of $G(m,n)$ then ${\cal K}$ is the curvature of the corresponding holomorphic solution of $G(n-m,n)$ and vice-versa. In particular, holomorphic solutions of $G(n-1,n)$ are obtained from  holomorphic solutions of $G(1,n)\cong{\mathbb C}P^{n-1}$.

Moreover, the special case $G(m,2m)$ enjoys an additional transposition invariance for all $m\geq1$. This invariance will be used, in particular,  to give a complete classification of holomorphic immersions of constant curvature of $S^2$ into $G(2,4)$.

\section{Constant curvature for holomorphic immersions in $G(m,n)$}

\subsection{General discussion and Veronese curves}

Let us recall that if we want holomorphic immersions in  $G(m,n)$ with  constant curvature ${\cal K}$,  the Lagrangian must take the form
\begin{equation}
{\cal L}\,= \frac{2}{\cal K} (1+\vert x\vert^2)^{-2},
\label{constc1}
\end{equation}
where $\vert x\vert^2=x_1^2+x_2^2=x_{+}x_{-}$. This implies that in  (\ref{em}) and  (\ref{em1}), the expression of $\det {\hat M}$ and $\det {\tilde M}$ must be proportional to $(1+\vert x\vert^2)^r$ where the positive integer $r$ is related to the curvature
\begin{equation}
{\cal K}=\frac{4}{r}.
\label{curvature}
\end{equation}
For $G(1,n)=\field{C}P^{n-1}$, we know that special solutions corresponding to the so-called Veronese curve $f^{(n)}: S^2 \to \field{C}P^{n-1}$ are given by
\begin{equation}
{f^{(n)}}
=\left(1, \sqrt{\left(\begin{array}{c}
n-1 \\
1 \end{array}\right)}\, x_+\,, \dots, 
\sqrt{\left(\begin{array}{c}
n-1 \\
r \end{array}\right)}\, {x_+}^r\,, \dots,
{x_+}^{n-1}
\right)^T\, ,
\label{veroneseseq}
\end{equation}
and have constant curvature  ${\cal K}=\frac{4}{n-1}$. Indeed, we have
$\det {\tilde M}= \vert { f^{(n)}}\vert^2=(1+\vert x\vert^2)^{n-1}$. We know that, up to gauge invariance, Veronese curves $f^{(k)}, \ k=1, \dots, n-1$, are the only solutions imbedded in $\field{C}P^{n-1}$ giving rise to holomorphic solutions${}^{7,9}$. 

We can generalize this result to the Veronese holomorphic curve in $G(m,n)$ which takes the form 
\begin{equation}
\tilde Z_V^{(m,n)}=\left(f^{(n)},\partial f^{(n)}, \dots, \partial^{m-1} f^{(n)}\right),
\label{tildeZvero}
\end{equation}
for which the curvature is now${}^{6}$
\begin{equation}
{\cal K}=\frac{4}{m(n-m)}.
\label{mincurv}
\end{equation}

We will conjecture in the following that this curve is the minimal constant curvature that we can find for holomorphic solutions.The corresponding integer $r$ in $\det {\tilde M}$, denoted $r_{\rm max}(m,n)$, is given by
 \begin{equation}
r_{\rm max}(m,n)=m(n-m)= {\rm dim} \ G(m,n).
 \label{rmn}
 \end{equation}
 
 Let us mention that this general result about the curvature of the Veronese curves may be easily recovered using our recent study${}^{6}$ of projectors associated with the Veronese sequence. Starting from the vector $\tilde Z_{V}^{(m,n)}$ given by (\ref{tildeZvero}), we construct a solution of the $G(m,n)$ model by the Gram-Schmidt orthogonalization process as usual and obtain
 \begin{equation}
Z_V^{(m,n)}=\left(\frac{f^{(n)}}{\vert f^{(n)}\vert}, \frac{P_+ f^{(n)}}{\vert P_+ f^{(n)}\vert}, \frac{P_+^2 f^{(n)}}{\vert P_+^2 f^{(n)}\vert}, \dots, \frac{P_+ ^{m-1} f^{(n)}}{\vert P_+ ^{m-1} f^{(n)}\vert}\right)
\label{tildeZvero1}
\end{equation}
where the operator $P_+$ is defined as 
\begin{eqnarray}
P_{+}: f \in \field{C}^N \rightarrow
P_{+}f=\partial_{+} f - \frac{f^{\dagger} \partial_{+} f }{\vert f \vert^2} f \,.
\label{operatorintermsoff}
\end{eqnarray}
A sequence of projectors $P_{k}$ could thus be constructed from these functions, {\it ie},
\begin{equation}
P_k:=\frac{{P}_+^k f \ {({P}_+^k f})^{\dagger}}
{\vert {P}_+^k f \vert^2},\,\,\, 
k =0,1,...,m-1\,,
\label{concproject}
\end{equation}

We have shown in ${}^{6}$ that the  curvature $\cal K$ is related to the following quantity 
 \begin{equation}
A\left(n,\sum_{i=0}^{m-1}P_i\right)=m(n-m),
\end{equation}
which is exactly the expected value (\ref{rmn}) of the power $r$. In the Macfarlane parametrization (\ref{part1}), we get  for the holomorphic Veronese curve in $G(m,n)$, the following expression,
\begin{equation}
({K_{V}^{(m,n)}})_{ij}=(-1)^{m-j}\left(\frac{m-j+1}{m-j+i}\right)\frac{\sqrt{{n-1 \choose i+m-1}}}{\sqrt{{n-1 \choose j-1}}}{i+m-1 \choose m}{m \choose j-1}x_{+}^{i-j+m},\label{KVmnij}
\end{equation}
for $i=1,...,n-m$ and $j=1,...,m$.


\medskip

\textbf{Theorem 2:} Holomorphic solutions with constant curvature of $G(m,n-1)$ generate holomorphic solutions with the same curvature of $G(m,n)$ by  embedding.

\textit{Proof:} Using the wedge product introduced in section 2 we get 
\begin{eqnarray}
 \det {\tilde M}(m,n)&=&\,\sum_{1\leq i_1<\cdots< i_m\leq n} \, \vert  {\cal A}_{i_1,i_2,\cdots i_m}\vert ^2\\
&=&\det  {\tilde M}(m,n-1)+\sum_{1\leq i_1<\cdots< i_{m-1}\leq n-1} \, \vert  {\cal A}_{i_1,i_2,\cdots,i_{m-1}n}\vert ^2.
\label{b}
\end{eqnarray}
Setting $(f_{i})_{n}=0$ for $i=1,...,m$ in the expression of  ${\cal A}_{i_1,i_2,\cdots,i_{m-1}n}$ given in (\ref{Agen}) , we obtain
\begin{equation}
 \det {\tilde M} (m,n)=\det {\tilde M}(m,n-1).\label{genimbed}
\end{equation}
 This concludes the proof.

\subsection{Admissible constant curvatures}

In the previous subsection, we have just seen that $G(m,n)$ admits a special type of holomorphic solutions with constant curvature such that $r=r(m,n)=m(n-m)$. Moreover, in different approaches${}^{1,2}$, a systematic classification of holomorphic solutions with constant curvatures has been given for $G(2,4)$ and $G(2,5)$ 
and it was shown that there exist such solutions for all integer values $r\leq m(n-m)$.

Based on these results and on our detailed investigations, we present here two our conjectures about  holomorphic solutions of $G(m,n)$ with constant curvature.

\medskip

{\bf Conjecture 1}: The maximal value of $r$ in the expression of $\det {\tilde M}$ ($\det {\hat M}$)
for which there exists a holomorphic solution of $G(m,n)$ of constant curvature is given by
$r_{\rm max}(m,n)=m(n-m)$. This solution corresponds to the Veronese holomorphic curve (\ref{tildeZvero}) and its
curvature is  minimal (\ref{mincurv}).

\medskip

This conjecture is clearly true for $G(1,n)=\field{C}P^{n-1}$. Due to the duality property, it is also true for $G(n-1,n)$. 

In the general case, since the expression of $\det {\tilde M}$ for $G(m,n)$ contains at most $N_m^n={n\choose m}=\frac{n!}{m!(n-m)!}$ terms, the maximal power $r$ is $N_m^n-1$. Since  $r_{\rm max}(m,n)\leq N_m^n-1$, in order to prove conjecture 1, we thus have to show that the values $r=r_{max}(m,n)+1,...,N_m^n-1$ are not possible. 

For example, for $G(2,n)$, we obtain $N_2^n=\frac{1}{2}n(n-1)$. This shows that the number of values of $r>r_{\rm max}(m,n)$ that have to be checked (and proved not to correspond to constant curvature solutions)  in order to prove our conjecture is thus given by
\begin{equation}
N_2^n-1-r_{\rm max}(2,n)=\frac{(n-3)(n-2)}{2}.
\end{equation}
In the next section, we prove our conjecture for $n=4,5$.

\medskip

{\bf Conjecture 2}: For $m$ fixed, holomorphic solutions with constant curvature in $G(m,n)$
can be constructed for all integer values of $r$ such that $1\leq r\leq r_{\rm max}(m,n)$. 

More precisely, the conjecture can be stated as follows: for $m$ fixed, all integers $r=r_{\rm max}(m,n)-k, \ \forall k=1, \dots, m-1$, give rise to holomorphic solutions of constant curvature in $G(m,n)$. 

Indeed, from Theorem 2, we know that $G(m,n)$ has holomorphic solutions for $r=1,...,r_{max}(m,n-1)$  by simple embedding from the holomorphic solutions of $G(m,n-1)$. The solution corresponding to $r=r_{max}(m,n)$ is given by the Veronese sequence (\ref{tildeZvero}). 

The conjecture is true also for $G(1,n)=\field{C}P^{n-1}$ and due to the duality property, it is also true for $G(n-1,n)$. 

In the next section, we discuss both conjectures for the case $G(2,n)$, for which some partial results  are already known${}^{1,2}$.

\section{Holomorphic solutions and constant curvatures of the $G(2,n)$ model for $n>3$}

We consider $n>3$ in the following developments. Indeed, for $G(2,3)$, the conjectures are true due to the duality with $\mathbb{C}P^2$. 

As mentioned before, constant curvature holomorphic solutions for the $G(2,n)$ case are obtained for $r=1,...,r_{max}(2,n-1)=2(n-3)$ by simple embeddings from the holomorphic solutions of $G(2,n-1)$. For  $r_{max}(2,n)= 2(n-2)$, we obtain the Veronese holomorphic solution from (\ref{tildeZvero}). It means that our conjectures will be proven if we show that  the values of $r$ given by 
\begin{equation}
r=r_{max}(2,n)+k, \ k=1, \dots, \frac{(n-3)(n-2)}{2},
\end{equation}
are not admissible (Conjecture 1) and that we can exhibit a solution for $r=2n-5$ (Conjecture 2).

Let us start with some general results on the explicit form of holomorphic solutions with constant curvature in $G(2,n)$. We consider the orthogonalized parametrization  (\ref{part1}) which can be written
\begin{equation}
{\hat Z}^{(2,n)} =\,\left(\begin{array}{c}
{\mathbb I}_2 \\
K^{(2,n)}
\end{array}\right)\,
\label{hatZ2}
\end{equation}
and such that
\begin{equation}
\det \hat M= \det ({\mathbb I}_2+ (K^{(2,n)})^{\dagger}K^{(2,n)}).
\label{detMgen}
\end{equation}

Gauge invariance (\ref{trans}) can now be used to obtain a canonical choice of $K^{(2,n)}$ and obtain a classification of solutions. In fact, previous approaches${}^{1,2}$ for the cases of the $G(2,4)$ and $G(2,5)$ models have explicitly used this gauge invariance. Due to the parametrization (\ref{hatZ2}), we can make the following transformations of the matrix $K^{(2,n)}$
\begin{equation}
K^{(2,n)}\to {K^{(2,n)}}'=V_0 K^{(2,n)} U_{0},
\label{Kprime}
\end{equation}
with constant matrices $V_0\in U(n-2)$ and $U_{0}\in U(2)$. 

As in recent papers${}^{1,2}$, the elements of the matrix $K^{(2,n)}$, can be chosen to be
\begin{equation}
{(K^{(2,n)})}_{i1}=\alpha_i x_+^{r_i},\qquad  {(K^{(2,n)})}_{i2}=\beta_i x_+^{s_i}, \ i=1,...,n-2.
\label{k2gen}
\end{equation}
Due to the  invariance property (\ref{Kprime}), we may choose $\beta_{1}, \alpha_{1},...,\alpha_{n-2}\in{\mathbb R}^{\geq0}$. Without loss of generality, we can also take $r_{1}=1\leq r_{2}\leq ...\leq r_{n-2}$ et $1\leq s_{1}$. Moreover, in order to obtain $\det \hat M$ as a function of $\vert x\vert^{2}$ only we have to satisfy the constraints
\begin{equation}
 s_{i}=r_{i}+s_{1}-1,\ i=1,...,n-2.
 \label{sj}
\end{equation}

We are now interested in the explicit expression for the determinant (\ref{detMgen}). This can be done using the wedge product introduced in section 2 (see (\ref{wedge1})) for the orthogonalized parametrization with $\hat Z^{(2,n)}=(\hat z)_{ik},  \ i=1,...,n,\ k=1,2$. We thus define
 \begin{equation}
 p_{ij}=\det\left(\begin{array}{cc}
  \hat z_{i1}&\hat z_{i2}\\
\hat z_{j1}&\hat z_{j2}
\end{array}\right), \ i,j=1,\dots ,n,\ i<j.
\end{equation}
The coordinates $p_{ij},   \ i,j=1, \dots ,n,\ i<j $ are, fact,  the so called Pl\"ucker coordinates${}^{2,10}$. They satisfy $p_{12}=1$ and the Pl\"ucker relations
\begin{equation}
 p_{ij}p_{kl}=p_{ik}p_{jl}-p_{jk}p_{il}\quad \Rightarrow \quad p_{ij}=p_{1i}p_{2j}-p_{1j}p_{2i},\label{Pluckrela}
\end{equation}
 for $i,j=3, \dots ,n, \, i<j$. From the explicit form of the elements of $K^{(2,n)}$ given in (\ref{k2gen}) we find that 
\begin{equation}
 p_{12}=1,\quad p_{1i}=\beta_{i-2}x_{+}^{r_{i-2}+s_1-1},\quad p_{2i}=-\alpha_{i-2}x_{+}^{r_{i-2}},\ i=3,\dots ,n
\end{equation}
 and 
\begin{equation}
 p_{ij}=\gamma_{i-2,j-2}x_{+}^{r_{i-2}+r_{j-2}+s_1-1}, \ i,j=3,\dots ,n, \ i<j
\end{equation}
with $\gamma_{i,j}=\alpha_{i}\beta_{j}-\alpha_{j}\beta_{i}$.  
 
The next step involves the verification that we can construct an embedding${}^{2,10}$ of $G(2,n)$ into $\mathbb{C}P^{N_2^n-1}$. Indeed, we introduce the map $\Phi_n:G(2,n)\rightarrow \mathbb{C}P^{N_2^n-1}$, where $N_2^n={\frac12} n(n-1)$ is the number of Pl\"ucker coordinates, such that
$\Phi_{n}(\hat Z^{(2,n)})=\mathcal{Z}_{N_2^n}$ with $\mathcal{Z}_{N_2^n}=(p_{ij})_{i<j}$ being a $N_2^n$ component vector. 

Then it is easy to prove that 
\begin{equation}
 \det {\hat M}^{(2,n)}=\det \hat Z^{(2,n)^{\dagger}}\hat Z^{(2,n)}=\mathcal{Z}_{N_2^n}^{\dagger}\mathcal{Z}_{N_2^n}.
 \label{detMplucker}
\end{equation}
For example, for $G(2,4)$, we have an embedding into $\mathbb{C}P^5$ and the map $\Phi_4$ is explicitly given by
\begin{eqnarray}
 \Phi_4(\hat Z^{(2,4)})=\mathcal{Z}_6&=&\left(
  1\ ,\ p_{23}\ ,\ p_{13}\ ,\ p_{24}\ ,\  p_{14}\ ,\ p_{34}\right)^T \nonumber \\
  &=&\left(
  1\ ,\ -\hat z_{31}\ ,\ \hat z_{32}\ ,\  -\hat z_{41}\ ,\  \hat z_{42}\ ,\ \hat z_{31}\hat z_{42}-\hat z_{32}\hat z_{41}
\right)^T.
\end{eqnarray}


Let us now present some general results for the $G(2,n)$ case dealing with extremal values of $r$ related to the quantity $N_2^n-1={\frac12} n(n-1)-1$.

\medskip

\noindent\textbf{Theorem 3:} In the Grassmannian manifold $G(2,n)$ for $n\geq6$, the value $r=N_2^n-1$ does not give rise to a holomorphic solution with constant curvature.\\
\textit{Proof:} In order to obtain the maximal power $r=N_2^n-1$ in the expression of $\mathcal{Z}_{N_2^n}$, all the Pl\"ucker coordinates $p_{ij}$ must be different from zero and the corresponding powers must be all different. The ordering of these powers implies that the two highest ones appear in $p_{n-1,n}$ and $p_{n-2,n}$. They are given by
\begin{equation}
 r_{n-3}+r_{n-2}+s_1-1=N_2^n-1,\quad r_{n-4}+r_{n-2}+s_1-1=N_2^n-2.
\end{equation}
We thus obtain $r_{n-3}=1+r_{n-4}$ and, consequently, 
\begin{equation}
 p_{1,n-1}=\beta_{n-3}x_{+}^{r_{n-3}+s_1-1}=\beta_{n-3}x_{+}^{r_{n-4}+s_1},\quad p_{3,n-2}=\gamma_{1,n-4}x_{+}^{r_{n-4}+s_1}.
 \label{ppn}
\end{equation}
This shows, however, that these two powers are the same which contradicts the assumption. This concludes the proof.

Let us mention that Theorem 3 is also valid for $n=4,\ 5$ but the proof is different and will be given in the next subsections.

\subsection{The case of $G(2,4)$}

In this case, we have
\begin{eqnarray}
\mathcal{Z}_6&=&\left(
  1\ ,\ p_{23}\ ,\ p_{13}\ ,\ p_{24}\ ,\  p_{14}\ ,\ p_{34}\right)^T \nonumber \\
  &=&\left(
  1\ ,\ -\alpha_{1} x_{+} \ ,\ \beta_{1} x_{+}^{s_1} \ ,\ -\alpha_{2} x_{+}^{r_{2}} \ ,\ \beta_{2} x_{+}^{r_{2}+s_{1}-1} 
   \ ,\ \gamma_{12} x_{+}^{r_{2}+s_{1}}
\right)^T.
\end{eqnarray}
Conjecture 1 will be proven if we show that the power $r=5$ is not allowed in the expression of  $\det  \hat M$ given by (\ref{detMplucker}). It is easy to prove, since we have  $r_{2}+s_{1}=5$ and, using the transposition invariance, $(r_{2},s_{1})=(3,2)$, we obtain the following constraints:
\begin{equation}
\alpha_{1}^2=\beta_{2}^2=5, \  \alpha_{2}^2= \beta_{1}^2=10
\end{equation}
which, together with
\begin{equation}
(\alpha_1 \beta_2-\alpha_2 \beta_1)^2=1,
\end{equation}
cannot be satisfied. For $r=4$, we get the solution corresponding to the Veronese sequence given by
\begin{equation}
 \mathcal{Z}_6=\left(1\ ,\ -2x_+\ ,\ \sqrt{3}x_+^2\ ,\ -\sqrt{3}x_+^2\ ,\ 2x_+^3\ ,\ x_+^4\right)^T.
\end{equation}

To prove Conjecture 2, we need to exhibit a solution for $r=3$. In this case, we can easily prove that $\gamma_{12}\neq0$ (or equivalently $p_{34}\neq 0$) and thus $r_{2}+s_{1}=3$. Again without loss of generality, we take $(r_{2},s_{1})=(2,1)$ and a holomorphic solution is obtained with
\begin{equation}
\mathcal{Z}_6=\left(1\ ,\ -\sqrt{\frac{8}{3}}x_+\ ,\ \frac{1}{\sqrt{3}}x_+\ ,\ -\sqrt{3}x_+^2\ ,\ 0\ ,\ -x_+^3\right)^T.
\label{kthree}
\end{equation}
The corresponding curvature is ${\cal{K}}=\frac{4}{3}$. 

Using, our approach, it is easy to reproduce the complete classification${}^{1,2}$ of holomorphic immersions of constant curvature in $G(2,4)$.

\subsection{The case of $G(2,5)$} 

In this case, the explicit expression of $\det {\hat M}^{(2,5)}$, as given in (\ref{detMplucker}), is
\begin{equation}
 \det {\hat M}^{(2,5)}=\mathcal{Z}^{\dagger}_{10}\mathcal{Z}_{10}=\sum_{1\leq i_1<i_2\leq5}\vert p_{i_1i_2}\vert^2,
 \label{detG25}
\end{equation}
where the ten component vector $\mathcal{Z}_{10}$ is given by
\begin{equation}
 \mathcal{Z}_{10}=(1\ ,\ p_{23}\ ,\ p_{13}\ ,\ p_{24}\ ,\ p_{14}\ ,\ p_{25}\ ,\ p_{15}\ ,\ p_{34}\ ,\ p_{35}\ ,\ p_{45})^T.\label{Z10}
\end{equation}

Let us recall that Conjecture 2 will be proven in this case if we obtain a solution of this type  for $r=5$. Such solutions have been already obtained${}^{2}$ but here we rederive two of them for completeness. The corresponding vectors $\mathcal{Z}_{10}$ in (\ref{Z10}) take the following forms:
\begin{eqnarray}
\mathcal{Z}_{10}&=&\left(1\ ,\ -\sqrt{5}x_+\ ,\ \sqrt{5}x_+^2\ ,\ -\sqrt{5}x_+^2\ ,\ \frac{7}{\sqrt{5}}x_+^3\ ,\ 0\ ,\ \frac{1}{\sqrt{5}}x_+^3\ ,\ 2x_+^4\ ,\ x_+^4\ ,\ x_+^5\right)^T,\\
\mathcal{Z}_{10}&=&\left(1\ ,\ -x_+\ ,\ 2x_+\ ,\ -\frac{1}{\sqrt{5}}x_+^2\ ,\ \frac{7}{\sqrt{5}}x_+^2\ ,\ 0\ ,\ \sqrt{5}x_+^3\ ,\ \sqrt{5}x_+^3\ ,\ \sqrt{5}x_+^4\ ,\ x_+^5\right)^T.
\end{eqnarray}
This representation of solutions shows an obvious symmetry between coefficients of both solutions. For $r=6$, we get the solution corresponding to the Veronese sequence
\begin{equation}
 \mathcal{Z}_{10}=\left(1\ ,\ \sqrt{6}x_+\ ,\ \sqrt{6}x_+^2\ ,\ 3x_+^2\ ,\ 4x_+^3\ ,\ 2x_+^3\ ,\ 3x_+^4\ ,\ -\sqrt{6}x_+^4\ ,\ -\sqrt{6}x_+^5\ ,\ -x_+^6\right)^T.
\end{equation}

It remains to prove Conjecture 1 which can be reformulated, in this special case, as the following proposition:
\medskip

\textbf{Proposition 1:} In $G(2,5)$, there are no holomorphic solutions with constant curvature corresponding to $r=9,\ 8,\ 7$.

\medskip
\textit{Proof:} In the expression of $\det {\hat M}^{(2,5)}$ given as (\ref{detG25}), we can see that
the highest powers of $|x_+|$ are $r=r_{3}+r_{2}+s_{1}-1$ and $r_{3}+s_{1}$ with the corresponding coefficients $\gamma_{23}$ and $\gamma_{13}$ in the Pl\"ucker coordinates.

(1) If $\gamma_{13}=\gamma_{23}=0$, we thus obtain $\gamma_{12}=0$ or $\alpha_{3}=0$. This means that the highest power one might expect is $r=6$. 

(2) If $\gamma_{23}=0$ and $\gamma_{13}\neq0$ (which is equivalent to the case $\gamma_{23}\neq0$ and $\gamma_{13}=0$), we first take $\beta_{3}=0$ then, either $\alpha_{3}=0$ or $\beta_{1}=0$, leaving us with a maximal power of $r=6$. 

We thus assume $\beta_{3}\neq 0$ and $\alpha_{1}=\frac{\alpha_{3}\beta_{1}}{\beta_{3}}$ and the maximal power one might expect is now $r_{3}+s_{1}=8$. In this case, we are lead to $(r_{2},s_{1})=(3,2)$ with the constraints $\alpha_{3}^{2}=\beta_{1}^{2}=28$ and
\begin{equation}
 \frac{\beta_{1}^{2}\alpha_{3}^{2}}{\beta_{3}^{2}}=\beta_{3}^{2}=8,
 \end{equation}
which cannot be simultaneously satisfied.

Let us now consider the case of $r_{3}+s_{1}=7$. We get the following not equivalent choices for $r_2$ and $s_1$: $(r_{2},s_{1})=(2,2)$, $(2,3)$, $(2,4)$ and $(3,2)$. It is trivial to check that they are not compatible with the constraint $\det M=(1+\vert x\vert^{2})^{7}$. 
 
For example, the case $(r_{2},s_{1})=(2,3)$ leads to $\frac{\alpha_{2}^{2}}{\alpha_{3}^{2}}=\frac{\beta_{2}^{2}}{7}=21$ and $\alpha_{3}^{2}=35-\beta_{2}^{2}$ which cannot be simultaneously satisfied.

(3) If $\gamma_{13}\neq0$ and $\gamma_{23}\neq0$, the maximal power that we can expect is $r=9$. In this case all the powers in (\ref{detG25}) must be different and all the Pl\"ucker coordinates are not zero. We thus obtain $r_{3}=8-s_{1}$ and $r_{2}=2$. With such a choice, we see that it is impossible to choose $s_{1}$ such that all the powers are different. 

Let us assume $r=8$. In this case, we have $r_{3}=7-s_{1}$ and $r_{2}=2$. We are thus led to three possible choices for $s_1$, namely,  $s_{1}=2,\ 3, \ 4$. It is straightforward to show that none of these cases is consistent with the constraint $\det M=(1+\vert x\vert^{2})^{8}$. 

Indeed, in the case $(r_{2},s_{1})=(2,2)$ we must have
\begin{equation}
 (\alpha_{2}\beta_{3}-\alpha_{3}\beta_{2})^{2}=1,\quad \vert\alpha_{3}\beta_{2}\vert=56,\quad \vert\alpha_{2}\beta_{3}\vert=\sqrt{28}\, l
\end{equation}
for $0\leq l\leq\sqrt{28}$ which are not compatible. In the case $s_{1}=3$, we show that $\alpha_{3}^{2}<0$ and in the case $s_{1}=4$, we deduce that $\beta_{3}^{2}<0$ which are both incompatible. Thus $r=8$ does not lead to holomorphic solutions with a constant curvature.

Let us now consider the most involved case of $r=7$. Again, we obtain $r_{3}=6-s_{1}$ and $r_{2}=2$. We are thus led to two choices for $s_1$, namely $s_{1}=2,\ 3$ subject to the constraint $\det M=(1+\vert x\vert^{2})^{7}$. Explicit but somewhat involved calculations show that both cases are incompatible (details in Appendix 1). Thus we have found that  $r=7$ does not lead to holomorphic solutions with constant curvatures. This completes the proof.

\medskip

As a final comment in this subsection let us mention that we know from Theorem 1 that the complete classification of such solutions for $G(2,5)$ gives rise to a complete classification of holomorphic solutions with constant curvatures for $G(3,5)$.

\bigskip

\section{Conclusions}

In this paper, we have found an explicit formula for the curvature of holomorphic immersions of $S^2$ into the general Grassmannian manifold $G(m,n)$. This formula has given us a way to look at general results concerning the classification of holomorphic maps of constant curvature. In particular, it allowed us to formulate two conjectures on the possible values of curvatures of such embeddings.  To test the validity of our conjectures we looked in detail on  the case of  $G(2,n)$ using Pl\"ucker coordinates and at an embedding into the well-known $\mathbb{C}P^N$ model. 

Both conjectures have been proven for the $G(2,4)$ and $G(2,5)$ models. In both cases, we obtained the full classifications of solutions and we did calculate the values of the corresponding constant curvatures. In the $G(2,4)$
case our results agree with the classification${}^{1}$ given by  Li and Yu and in the $G(2,5)$ 
case we have reproduced the  classification presented in ${}^{2}$. 

So far we have not succeeded in proving  our conjectures  even for the $G(2,n)$ case.
Indeed, in order to prove Conjecture 2, it remains to demonstrate the case of $r=2n-5$, which is not easy  for $n\geq 6$.
However we believe it to be correct and  Appendix 2 we produce a numerical proof of the existence of holomorphic solutions with $r=7$ in the $G(2,6)$ case.

In order to prove Conjecture 1, we have to make a deep analysis of the nonlinear constraints on the 
Pl\"ucker coordinates.

\section{Appendix 1: The case of $r=7$ in $G(2,5)$}

To complete the proof of Proposition 1, we need to show that, in the case of $\gamma_{13}\neq0$ and $\gamma_{23}\neq0$, the value $r=7$ is not admissible. We have two possible candidates corresponding to $(r_2,s_1)=(2,2)$ and $(r_2,s_1)=(2,3)$. In both cases, we have to impose $\alpha_1=\sqrt{7}$.

For $(r_2,s_1)=(2,2)$, we have to satisfy the additional constraints
\begin{equation}
 \gamma_{13}^2=7\,\gamma_{23}^2=7, \, \, \alpha_2=\sqrt{21}\sin\theta, \, \, \beta_1=\sqrt{21}\cos\theta, \, \, \beta_3=\sqrt{21}e^{i\psi}, \, \, \beta_2=\sqrt{35}e^{i\eta},
\end{equation}
where $0<\theta<\frac{\pi}{2}$ (using gauge invariance). We, then, deduce that
\begin{equation}
 \alpha_3=\frac{e^{i\lambda_1}+21e^{i\psi}\sin\theta}{\sqrt{35}e^{i\eta}}=\frac{e^{i\lambda_2}+\sqrt{21}e^{i\psi}}{\sqrt{3}\cos\theta},
\end{equation}
from which we calculate $\alpha_3^2$ and end up with  the following equation to satisfy
\begin{equation}
 21\cos^2\theta(1+21^2\sin^2\theta+42\sin\theta\cos(\psi-\lambda_1))=5390+490\sqrt{21}\cos(\psi-\lambda_2).
\end{equation}
We, then, maximize the left hand side by imposing $\cos(\psi-\lambda_1)=1$ (note that $0<\theta<\frac{\pi}{2}$) which becomes a function of one variable $\theta$. This can, in turn, be maximized but lead to the following impossible constraint
\begin{equation}
 \cos(\psi-\lambda_2)\approx-1.22454.
\end{equation}

For the $(r_2,s_1)=(2,3)$ case, we get the following constraints on the coefficients
\begin{equation}
 \gamma_{13}^2=7,\, \, \alpha_3=\sqrt{35}\cos\theta, \, \, \beta_1=\sqrt{35}\sin\theta, \, \, \alpha_2=\sqrt{21}, \, \, \beta_2=\sqrt{35}e^{i\psi},
\end{equation}
where, by gauge invariance, $0<\theta<\frac{\pi}{2}$. We, thus, deduce that
\begin{equation}
 \beta_3=e^{i\lambda_2}+5\sqrt{7}\sin\theta\cos\theta\,\,\Rightarrow\,\, \beta_3^2=1+175\sin^2\theta\cos^2\theta+10\sqrt{7}\sin\theta\cos\theta\cos\lambda_2.
\end{equation}
Furthermore, we may calculate $\gamma_{12}^2$ which has to satisfy
\begin{equation}
 \gamma_{12}^2=35(7+21\sin^2\theta-2\sqrt{7}\sqrt{21}\sin\theta\cos\psi),
\end{equation}
and thus we are left with the equation $\gamma_{12}^2+\beta_3^2=21$ to satisfy. We then minimized the left hand side by imposing $\cos\psi=-\cos\lambda_2=1$. We thus get two possible values of $\theta$: $\theta\approx0.0901758$ and $\theta\approx1.49397$ and we obtain
\begin{equation}
 \gamma_{12}^2+\beta_3^2\vert_{\theta\approx0.0901758}\approx179.642>21, \, \, \, \gamma_{12}^2+\beta_3^2\vert_{\theta\approx1.49397}\approx129.491>21,
\end{equation}
which contradicts $\gamma_{12}^2+\beta_3^2=21$. This concludes the proof of Proposition 1.

\section{Appendix 2: Numerical evidence of $r=7$ in $G(2,6)$}
In order to prove the conjecture 2 in the $G(2,6)$ case, we must find a holomorphic solution of constant curvature $\mathcal{K}=\frac{4}{7}$. To do this we consider the following generalization of solutions $\mathcal{Z}_6$ and $\mathcal{Z}_{10}$ by considering the following Pl\"ucker coordinates of $\mathcal{Z}_{15}$:
\begin{eqnarray*}
p_{23}&=&-\sqrt{7}\,x_+,\, \, \, \, \ p_{13}=\beta_1\, x_+^2,\, \, \, \, p_{24}=-\alpha_2\, x_+^2,\, \, \, \, p_{14}=\beta_2\,x_+^3,\\ p_{25}&=&-\frac{1}{\beta_4}\,x_+^3,\, \, \, \, p_{15}=\beta_3\, x_+^4,\, \, \, \, p_{26}=0,\, \, \, \, p_{16}=\beta_4\, x_+^4.
\end{eqnarray*}
 This choice implies that the Pl\"ucker coordinate $p_{56}$, which contains the highest powers of $x_+$ in $\mathcal{Z}_{15}$, is such that $p_{56}=x_+^7$. The remaining coordinates then become
\begin{eqnarray*}
 p_{34}&=&(\sqrt{7}\beta_2-\alpha_2\beta_1)\,x_+^4,\, \, \, \, p_{35}=(\sqrt{7}\beta_3-\frac{\beta_1}{\beta_4})\,x_+^5,\\ p_{36}&=&\sqrt{7}\beta_4\, x_+^5,\, \, \, \, p_{45}=(\alpha_2\beta_3-\frac{\beta_2}{\beta_4})\,x_+^6,\, \, \, \, p_{46}=\alpha_2\beta_4\, x_+^6,
\end{eqnarray*}
which most satisfy the constraints
\begin{eqnarray}
 p_{13}^2+p_{24}^2&=&p_{35}^2+p_{36}^2=21, \, \, \, \, \, \, \, p_{45}^2+p_{46}^2=7,\\
p_{14}^2+p_{25}^2&=&p_{15}^2+p_{16}^2+p_{34}^2=35.
\end{eqnarray}

 This set of quadratic constraints on the Pl\"ucker coordinates justified the use of a mathematical program (we used \textit{Mathematica} 8.) A symbolic solution of these constraints was found by the program. We then, convinced ourselves that the symbolic solutions for the Pl\"ucker coordinates were the good ones by taking numerical approximations of them. Indeed, we verified that the numerical approximations were such that $\mathcal{Z}_{15}^{\dagger}\mathcal{Z}_{15}\approx(1+\vert x\vert^2)^7$. Even though this is a numerical verification of the compatibility condition, the fact that \textit{Mathematica} 8 has originally produced a symbolic solution is an evidence of the correctness of conjecture 2 in $G(2,6)$. Here, we give one approximate solution  to our problem
\begin{equation}
 \beta_1\approx-0.4907042,\,\,\, \beta_2\approx2.8363697,\,\,\, \beta_3\approx2.6842282,\,\,\, \beta_4\approx-0.1926106,\,\,\, \alpha_2\approx-4.5562275.
\end{equation}

\section*{Acknowledgments}
This work reported here has been supported in part by a research grant
from NSERC of Canada. LD also acknowledges a FQRNT
fellowship.


\end{document}